\documentclass[aps,prl,twocolumn,groupedaddress,superscriptaddress,showpacs]{revtex4}

\usepackage{graphicx}
\usepackage{mathrsfs}

\begin{document} 

\title{Onset of ferromagnetism in low-doped Ga$_{\rm 1-x}$Mn$_{\rm x}$As} 
\author{B. L. Sheu}
\affiliation{Physics Department and Materials Research Institute, Pennsylvania State University, University Park, PA 16802}
\author{R. C. Myers}
\affiliation{Center for Spintronics and Quantum Computation, University of California, Santa Barbara, CA 93106}
\author{J.-M. Tang}
\affiliation{Optical Science and Technology Center and Department of Physics and Astronomy, University of Iowa, Iowa City, IA 52242}
\author{N. Samarth}
\affiliation{Physics Department and Materials Research Institute, Pennsylvania State University, University Park, PA 16802}
\author{D. D. Awschalom}
\affiliation{Center for Spintronics and Quantum Computation, University of California, Santa Barbara, CA 93106}
\author{P. Schiffer}
\affiliation{Physics Department and Materials Research Institute, Pennsylvania State University, University Park, PA 16802}
\author{M. E. Flatt\'e}
\affiliation{Optical Science and Technology Center and Department of Physics and Astronomy, University of Iowa, Iowa City, IA 52242}

\begin{abstract}
We develop a quantitatively predictive theory for impurity-band ferromagnetism in the low-doping regime of Ga$_{\rm 1-x}$Mn$_{\rm x}$As and compare with experimental measurements of a series of samples whose compositions span the transition from paramagnetic insulating to ferromagnetic conducting behavior. The theoretical Curie temperatures depend sensitively on the local fluctuations in the Mn-hole binding energy, which originates from disorder in the Mn distribution as well as the presence of As antisite defects.   The experimentally-determined hopping energy at the Curie temperature is roughly constant over a series of samples whose  conductivities vary more than $10^4$ and whose hole concentrations vary more than $10^2$. Thus in this regime the hopping energy is an excellent predictor of the Curie temperature for a sample, in agreement with the theory. 
\end{abstract}


\maketitle 

The III-V ferromagnetic semiconductor Ga$_{\rm 1-x}$Mn$_{\rm x}$As has been the focus of intense interest due to its potential incorporation in proof-of-concept spintronic devices\cite{Awschalom2002,Samarth2004,MacDonald2005,Awschalom2007}. Most studies have focused on relatively high Mn concentrations (x $\sim$ 0.03 - 0.08)\cite{Ohno1996,Ku2003,Chiba2003b,Nazmul2005} for which the Curie temperature can reach values of well over 100 K.  
Theoretical treatments of the Curie temperature in the low doping regime\cite{Bhatt-PRL-2001,Kaminski2002,Erwin2002,Fiete2003} have focused  on the  Mn dopants in the sample, including their concentration, distribution, compensation, and whether they are substitutional or interstitial. Those that treated variable-range hopping (VRH)\cite{Erwin2002} considered the Efros-Shklovskii regime\cite{Efros1975}, where the hopping energy arises from the Coulomb interaction, and found an even {\it stronger} dependence of the Curie temperature on hole doping than in the higher-doped regime.  Recent experimental measurements on low-doped Ga$_{\rm 1-x}$Mn$_{\rm x}$As, however, show the dependence of the Curie temperature on hole doping to be negligible, or even nonexistent\cite{Myers2006}, such that samples with the same measured hole density and Mn density may differ in Curie temperature by a factor of 2. This suggests the proper VRH regime is Mott VRH\cite{Mott1969,Apsley1974,VanEsch1997}, in which hopping energies are independent of the hole density.

Here we present a quantitative theory of the Curie temperature of Ga$_{\rm 1-x}$Mn$_{\rm x}$As in the Mott VRH regime which is built on a microscopic model for the effect of disorder on spin-spin interactions. We also present measurements of the Mn concentration, hole concentration, conductivity, hopping energy, and Curie temperatures in a series of samples prepared via a 'combinatorial' approach that exploits a spatial gradient in the Ga:As flux ratio on non-rotated substrates during MBE growth.  We use our theory to estimate the hopping energy from the density of substitutional Mn (ionized or neutral), and find that it is comparable to the ferromagnetic interaction energy between two spins. We also present self-consistent calculations of the Curie temperature in the Mott VRH regime, and we find the Curie temperature to be controlled primarily by the hopping energy, in agreement with the experimental findings.

\begin{figure}
\includegraphics[width=\columnwidth]{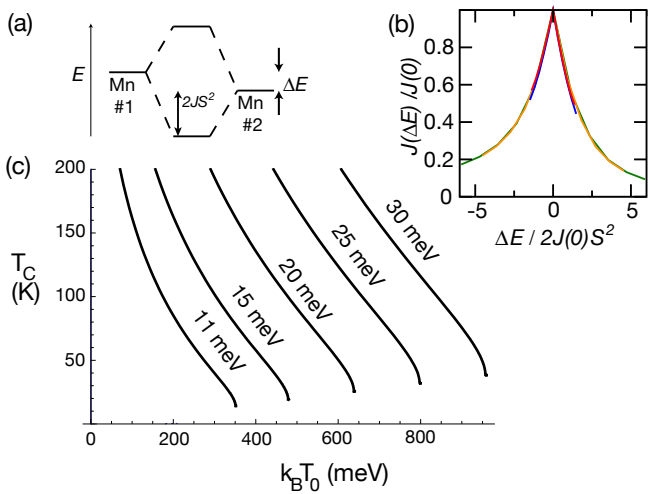}
\caption{(color online) (a) Model taking into account an energy difference $\Delta E$ between the two Mn hole states. (b) Ferromagnetic interaction energy for Mn-Mn pairs separated by  17\AA. Line between the two Mn is along the [221] direction, with sample magnetization along [001] (red) or [100] (blue), or along the [100] direction, with sample magnetization along [001] (orange) or [100] (green). (c) $T_c$ as a function of $T_0$ for several averaged values of $2J(0)S^2$ for 17\AA\ pairs.}
\label{mn-mn-interaction}
\end{figure}

We begin by describing our theoretical model and results.  The typical hopping distance and hopping energy play an important role in Mott VRH, for sites that are spatially close typically have energy differences too large to permit rapid hopping, whereas those that are spatially far away have small hopping matrix elements. The hopping energy $W$ is the typical energy difference between two sites separated by the most likely hopping distance $\ell$ of a carrier, and can be directly extracted from the temperature dependence of the conductivity, which follows a $T^{1/4}$ dependence. As noted in Ref.~\cite{Erwin2002}, similar quantities influence ferromagnetic interactions between Mn dopants. In our model, however, $W$ and $\ell$ are determined by the availability of states, rather than interactions between holes. Shown in Figure~\ref{mn-mn-interaction}(a) is a schematic double exchange model for two interacting Mn with parallel spins, including a site energy difference $\Delta E$. As in typical two-site models of ferromagnetism\cite{Anderson1963}, for parallel spins the hole bound state associated with Mn \#1 hybridizes with that of Mn \#2, and one molecular state has lower energy than either isolated Mn state. If a single hole is present per pair then parallel alignment of the Mn spins is preferred to antiparallel alignment (in which the states cannot hybridize). Bound hole states at different energies $\Delta E$ hybridize more weakly  and lead to a reduced ferromagnetic coupling as shown in Fig.~\ref{mn-mn-interaction}(b).

We calculate the interactions between two Mn in the presence of disorder, $H=JS_1\cdot S_2$, in this dilute carrier regime. Shown in Fig.~\ref{mn-mn-interaction} are calculated values for $J(\Delta E)/J(0)$ as a function of $\Delta E$ obtained from a multiband tight-binding theory\cite{Tang2004} that has successfully predicted the interaction energies of Mn-Mn pairs measured by STM\cite{Kitchen2006}.  $J(0)$ depends on the specific Mn-Mn pair, yet the dependence of $J(\Delta E)$ on $\Delta E$ is reproduced by an analytic expression from the model in Fig.~\ref{mn-mn-interaction}(a),
\begin{equation}
J(\Delta E)/J(0) = ((\Delta E/4J(0)S^2)^2 + 1)^{1/2} - |\Delta E|/4J(0)S^2. \label{JofE}
\end{equation}
Plotting the results for Mn-Mn pairs separated by 17\AA\ in scaled units collapses all the curves onto the universal shape shown in  Fig.~\ref{mn-mn-interaction}(b). At a distance of 17\AA\ the calculated interaction energy $2J(0)S^2$ varies from 8.5~meV to 36~meV, depending on the pair geometry. For other distances the average interaction energy can be estimated using the approximate decay length of the acceptor wave function, $\alpha^{-1} = 13$\AA.\cite{Tang2005}

To evaluate the effect of hopping transport on the ferromagnetic interaction we assume the most important pairs for carrier-mediated ferromagnetism are those providing the dominant contribution to the conductivity, and use the hopping energy 
\begin{equation}
W(T)=[\Gamma(5/4)/2] k_B T_0^{1/4}T^{3/4}\label{hopping}
\end{equation}
 for $\Delta E$.  From the results of Fig.~\ref{mn-mn-interaction}, once $W$ becomes of the order of $2J(0)S^2$ then the ferromagnetic interaction will be significantly reduced. We follow Ref.~\onlinecite{Apsley1974} to evaluate $\ell$ and $T_0$. Calculating the hybridization between Mn sites\cite{Tang2004,Kitchen2006} separated by the typical inter-Mn distance for $x=0.015$, $15$\AA, yields an estimate of 200~meV for the bandwidth, similar to that found in Ref.~\onlinecite{VanEsch1997}. There are $3$ states per Mn, and thus $k_BT_0\sim 700$~meV. The Curie temperature can be found by solving 
\begin{eqnarray}
\ell(T_c)&=&\alpha^{-1}\Gamma(5/4)(T_0/T_c)^{1/4}/2,\label{ell}\\
 T_c &=& S(S+1) J_\ell(W(T_c))/3k_B.\label{tc}
 \end{eqnarray}
simultaneously with Eqs.~(\ref{JofE})-(\ref{hopping}). $T_c$'s are shown in Fig.~\ref{mn-mn-interaction}(c) for several values of $2J(0)S^2$ for $17$\AA\ pairs.

We now compare the results of this model with the magnetic and electronic properties of a series of samples within which the electronic properties were continuously varied as described in detail in Ref.~\cite{Myers2006}.  We grew 100~nm thick Ga$_{\rm 1-x}$Mn$_{\rm x}$As epilayers on top of a 300~nm GaAs buffer on semi-insulating (001) GaAs wafers for x = 0.0075, 0.01, 0.0125, and 0.015, where the quoted Mn concentration values are calibrated from reflection high-energy electron diffraction (RHEED) measurements and secondary ion mass spectroscopy (SIMS).  Due to the arsenic flux gradient\cite{Myers2006}, the carrier density varied across each wafer by at least two orders of magnitude from compensation by As antisites. We cleaved three long strips side-by-side ($\sim 4$~mm in width) from the center portion of each wafer along the As-gradient direction.  Each strip was then cleaved into 15 sample pieces ($\sim 3$~mm $\times$ 4~mm) for transport, magnetization, and SIMS measurements of samples with varying As:Ga stoichiometry.

We measured magnetization with a commercial superconducting quantum interference device (SQUID) magnetometer (Quantum Design MPMS), whereas resistivity and Hall effect measurements were performed in commercial cryostats (Quantum Design PPMS) using four-probe techniques with external electronics to measure resistances up to $> 1$G$\Omega$.  We determined the existence of ferromagnetism, and  $T_c$, from temperature-dependent magnetization data, taken on warming in a 50~Oe field after initial cooling from room temperature to 2~K in a 1~T magnetic field.  Measurements of $M(H)$ hysteresis loops confirmed the existence of ferromagnetism. We used annealed indium contacts on lithographically patterned Ga$_{\rm 1-x}$Mn$_{\rm x}$As hall bars for transport measurements, and the contacts were confirmed to be ohmic in the source current range of $10-1000$~nA and $T = 2-300$K.

\begin{figure}
\includegraphics[width=\columnwidth]{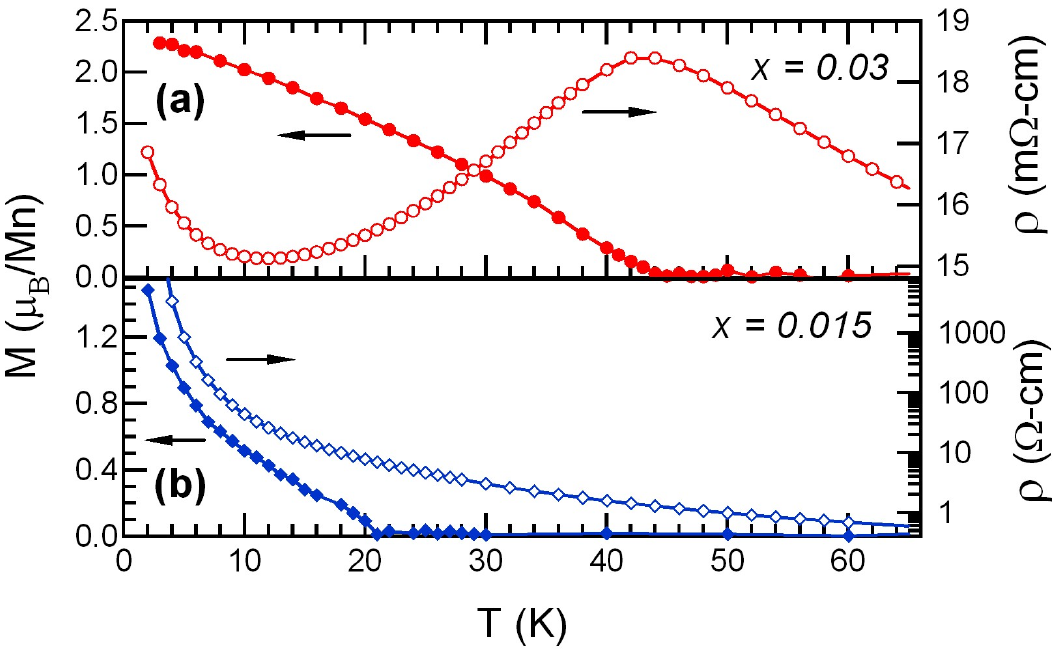}
\caption{(color online) Temperature dependence of the resistivity (open symbols) and the magnetization (closed symbols) along the [100] crystalline direction of as-grown (Ga,Mn)As epilayers grown in rotated mode for x = 0.03 and 0.015. }
\label{fig2}
\end{figure}

Figure~\ref{fig2} shows the temperature dependent resistivity $\rho(T)$ and the magnetization $M(T)$ along the [100] crystalline direction of Ga$_{\rm 1-x}$Mn$_{\rm x}$As epilayers synthesized using the conventional growth technique (rotating the substrate for uniformity). The peak observed near $T_c$ in the $\rho(T)$ data of Fig.~\ref{fig2}(a) is typical behavior for metallic high-$T_c$ Ga$_{\rm 1-x}$Mn$_{\rm x}$As.\cite{Awschalom2002,Samarth2004}  For smaller values of x, near the onset of ferromagnetism, insulating behavior ($d\rho/dT < 0$) is observed at all measured temperatures, as shown in Fig.~\ref{fig2}(b), and there is effectively no feature near $T_c$. These data clearly demonstrate a ferromagnetic insulating state in Ga$_{\rm 1-x}$Mn$_{\rm x}$As.

\begin{figure}
\includegraphics[width=\columnwidth]{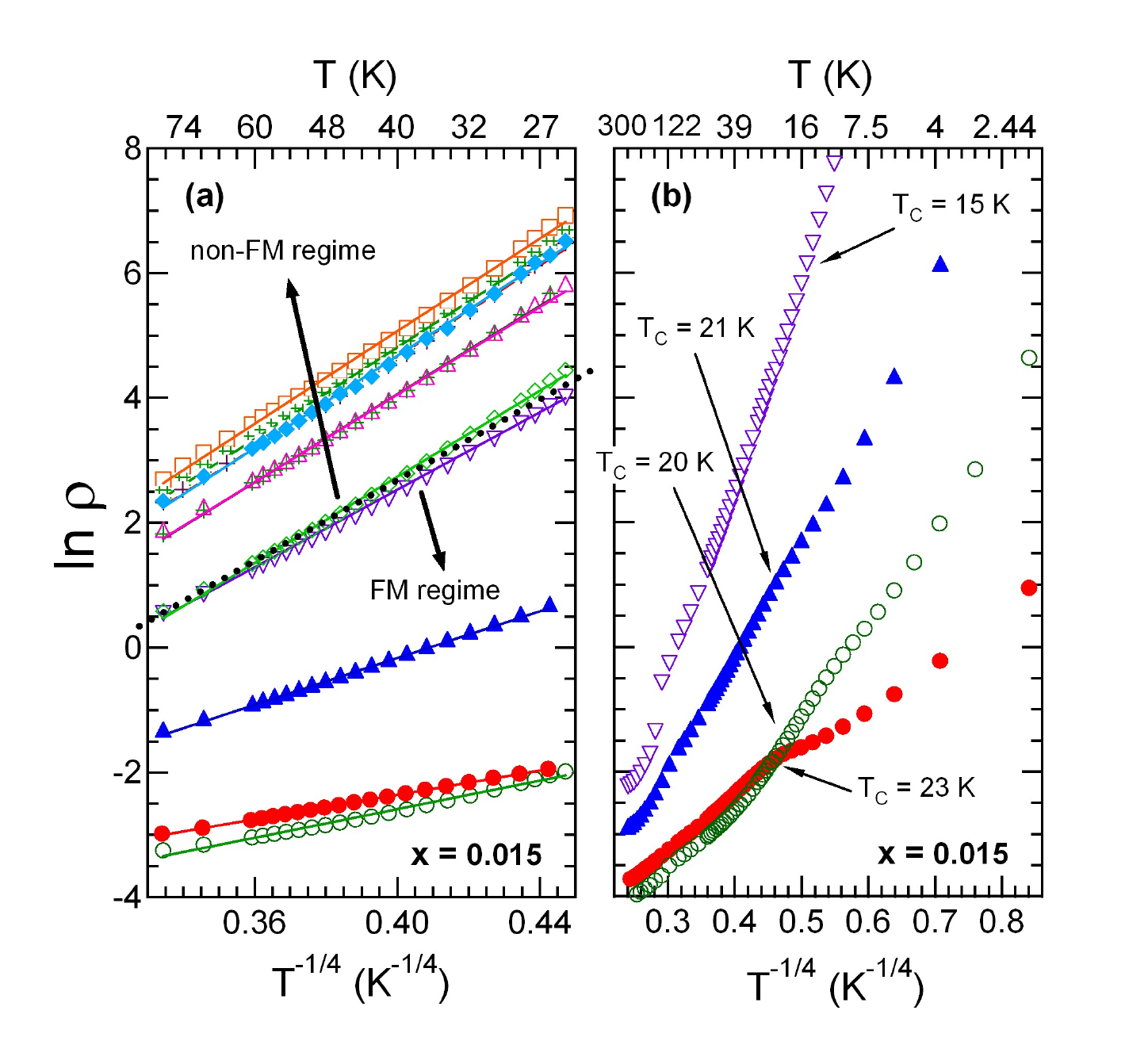}
\caption{(color online)  Resistivity from 25 K to 80 K for x = 0.015 non-rotated samples.  The straight lines of the same color are the corresponding fits performed in this temperature range as discussed in the text. (b) The full temperature range T = 2 - 300 K for all FM samples from the x = 0.015 series, where a noticeable deflection appears near $T_c$ and separates the curve into two segments.}
\label{fig3}
\end{figure}

We interpret the temperature dependence of the resistivity through the VRH expression $\rho = \rho_0\exp(T_0/T)^{1/4}$, although we caution that this analysis can only be applied to our data over a limited temperature range. $W(T_c)$ is then calculated from $T_0$ via Eq.~(\ref{hopping}).  Figure~\ref{fig3}(a) shows ${\rm ln}\rho(T)$  as a function of $T^{-1/4}$ between 80~K and 25~K (i.e. above $T_c$) for all of the pieces from the x = 0.015 non-rotated sample.  The solid lines are linear fits performed in this temperature range, and the single dashed line in the center of the graph separates the ferromagnetic (FM) and non-ferromagnetic (non-FM) regime (as determined from magnetization measurements). Similar trends and quality of fit were observed in the transport data sets for the other Mn concentrations studied. Figure~\ref{fig3}(b) shows the full temperature range of $2 - 300$~K for all FM samples for the x = 0.015 series. A noticeable deflection appears near $T_c$, suggesting that the magnetic and transport properties are still connected even in this low doping regime (although a similar deflection has been reported in a non-magnetic doped GaAs system)\cite{Redfield1973}.

\begin{figure}
\includegraphics[width=\columnwidth]{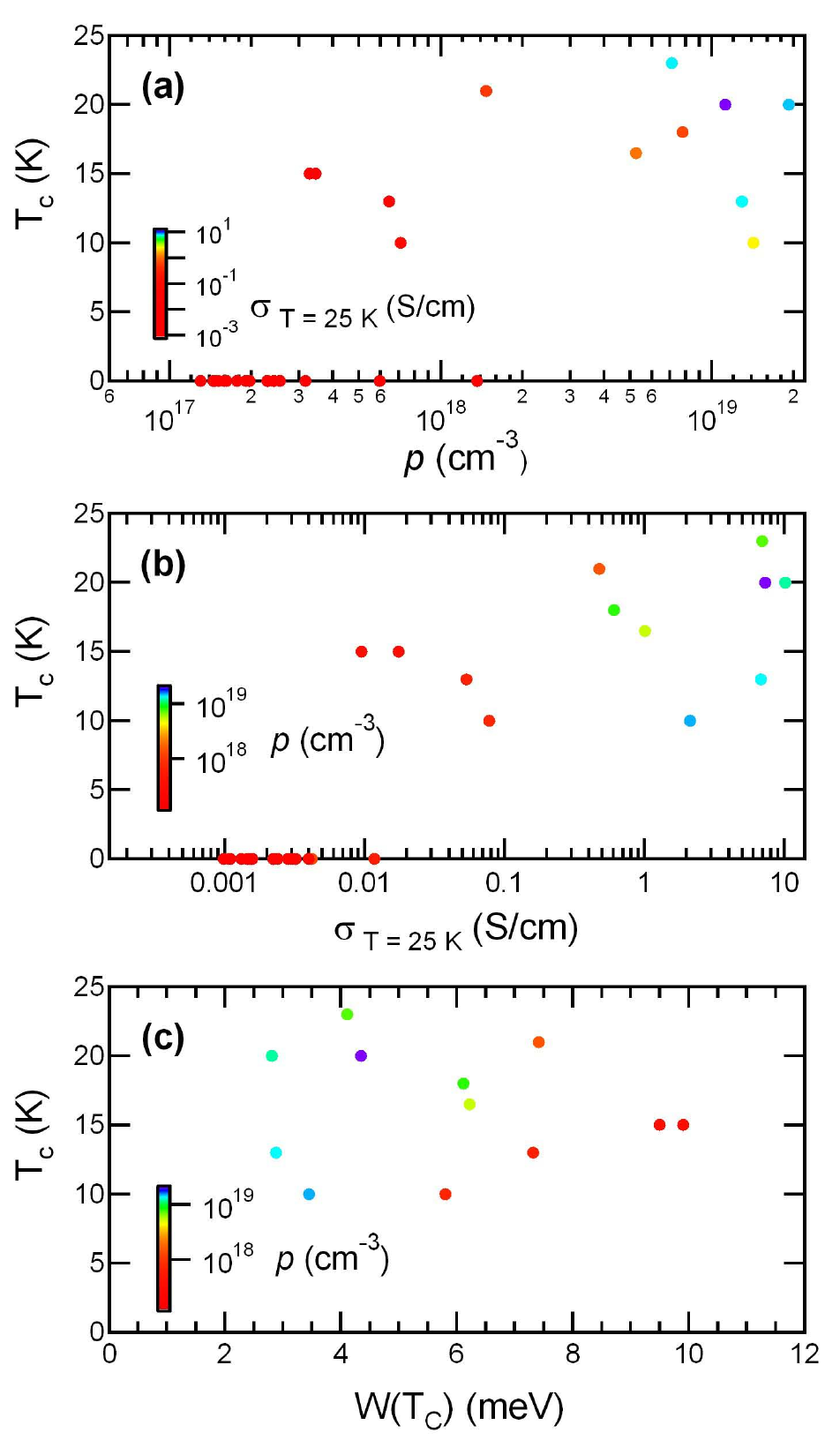}
\caption{(color online)  Curie temperature of non-rotated Ga$_{\rm 1-x}$Mn$_{\rm x}$As samples with x = 0.0075, 0.01, 0.0125, and 0.015 plotted as a function of (a) carrier concentration at 300 K (b) conductivity at 25 K and (c) hopping energy at $T_c$. }
\label{fig4}
\end{figure}

We plot the $T_c$ of all of our non-rotated samples in Figure~\ref{fig4} as a function of carrier concentration and conductivity at T = 25 K, and hopping energy at the Curie temperature.  The carrier concentration is derived from Hall effect measurements between $-2$~T and 2~T taken at 300 K due to the high resistance at low temperatures.   Every data point in Fig.~\ref{fig4} represents an individual cleaved piece from the wafer strips as described before. Although not explicitly shown, there are no conductivity data points for $W(T = 25K) > 21$~meV since the sample resistance at $25$~K is too large to measure with our apparatus. Despite the exceptionally broad distribution of conductivities (over four orders of magnitude) and hole concentrations (over two), all of the ferromagnetic materials have $W(T_c)$ between 3 and 10~meV.

The value of $\ell$ extracted for the ferromagnetic samples is 22\AA\ (as $\ell$ depends on $T_c^{1/4}$, there is little variation in $\ell$ for this range of samples). The estimated value of $2J_\ell(0)S^2$ for this distance is then $8$~meV, which is within a factor of two of the measured hopping energy at the Curie temperature for all the samples, in agreement with the theory. The bandwidth, which influences $T_0$, does depend on the Mn concentration, but for a change of Mn concentration of a factor of $2$ both $W$ and $\ell$ will only change by $2^{1/4}$, which for the measurements made here has a negligible effect on $T_c$. The absence of ferromagnetic samples of $T_c<10$K is consistent with the solutions shown in Fig.~\ref{mn-mn-interaction}(c), where a discontinuous drop from a finite $T_c$ to non-ferromagnetic material is seen.

The high carrier density limit ($\sim 10^{19}$~cm$^{-3}$) of these low Mn-doped samples just reaches the lower limit of metallic Ga$_{\rm 1-x}$Mn$_{\rm x}$As considered in typical theoretical calculations. Compared to other Ga$_{\rm 1-x}$Mn$_{\rm x}$As samples with a similar Mn-doping level, the resistivity of our ferromagnetic samples is several orders of magnitude lower, which is possibly due to an optimized control of the arsenic and the reduction of As-antisites. This reduced disorder will lead to a smaller bandwidth, and thus a smaller $T_0$, which (see Fig.~\ref{mn-mn-interaction}(c)) permits the materials to be ferromagnetic. Our data show that the carrier concentration has a weak effect on the onset of ferromagnetism in this limit. Instead, the strong correlation of the cutoff in ferromagnetism with critical values of the electric conductivity and the hopping energy suggest that the interplay between disorder and Mn-Mn interactions found in Eqs.~(\ref{JofE})-(\ref{tc}), and shown in Fig.~\ref{mn-mn-interaction}(c), dominate the transition.  The data in Fig.~\ref{fig4}, encompassing samples over a broad range of As:Ga stoichiometry, indicate that the key experimental quantity determining the Curie temperature is the hopping energy (or, equivalently, $T_0$), regardless of the microscopic details of the disorder. We even find similar results for Ga-rich samples (not shown here), which are structurally much less homogeneous\cite{Myers2006} than the As-rich samples described here. As further support for this view, samples with $W(T=25K)>15$~meV were found to be non-ferromagnetic, whereas all samples with $W(T=25K)<15$~meV were ferromagnetic.

These results currently apply to materials which are not at the upper extreme of Curie temperatures. However, magnetic semiconductors in the low-doped regime\cite{Myers2005} have longer nonequilibrium carrier and spin lifetimes, perhaps making them more suitable for magneto-optoelectronic applications. The theory presented here might also be extended to higher hole densities. Screening by holes of disorder potentials would lead to a narrower bandwidth and thus higher $T_c$, so the theory presented here would become sensitive to the hole density at higher concentrations. If it is possible to extend the control of As antisite concentration pioneered in Ref.~\onlinecite{Myers2006} to higher Mn densities perhaps still higher Curie temperatures may be achieved in very high conductivity samples. 

Our studies of the onset of ferromagnetism in a magnetic semiconductor have revealed a mechanism for ferromagnetism in the Mott variable-range-hopping regime that is not well described by the commonly-used mean-field theory based on the local carrier density. Even though carriers are required for formation of the ferromagnetic state, the Curie temperature is almost entirely insensitive to the hole density in this regime. The total magnetization of the material, however, will be affected as the hole density is reduced. Increased disorder leads to a larger bandwidth and a lower Curie temperature, and eventually to the discontinuous disappearance of the ferromagnetic state.    Measurements on samples with the same experimental Mn concentration and hole concentration, but very different hopping energies, also indicate a strong correlation between hopping energy and Curie temperature, and a weak correlation with hole density, along the lines of the theory. 

We would like to thank E. Dagotto and A. Moreo for valuable comments. We acknowledge ONR Grant Nos. N00014-06-1-0428 and N00014-05-1-0136, the NSF NNIN node at PSU, and NSF Grant Nos. DMR 0401486, DMR-0305238, and DMR-0305223. 

\bibliography{mn-vrh}

\end{document}